\documentclass[conference,twocolumn]{IEEEtran}
\IEEEoverridecommandlockouts

\usepackage[subtle,indent=normal,mathspacing=normal,wordspacing=normal,title=normal,charwidths=normal]{savetrees}

\usepackage{balance}

\usepackage[dvipsnames]{xcolor}

\usepackage[cmex10]{amsmath}
\usepackage{amssymb,amsfonts,amssymb,amsthm}
\usepackage{stmaryrd}
\usepackage{mathtools}
\usepackage[T1]{fontenc}
\usepackage{cite}
\usepackage{graphicx}

\usepackage{colortbl}
\usepackage[font=footnotesize]{subcaption}

\usepackage[hidelinks]{hyperref}
\usepackage{crossreftools}
\usepackage{cleveref}
\Crefname{figure}{Fig.}{Figs.}

\usepackage{pgfplots}
\usepackage{pgfplotstable}
\usepackage{tikz}
\usetikzlibrary{calc,decorations.pathreplacing}

\usepgfplotslibrary{fillbetween}

\DeclareCaptionLabelSeparator{periodspace}{.\quad}
\def\BibTeX{{\rm B\kern-.05em{\sc i\kern-.025em b}\kern-.08em
T\kern-.1667em\lower.7ex\hbox{E}\kern-.125emX}}

\newtheorem{theorem}{Theorem}

\makeatletter
\newcommand{\removelatexerror}{\let\@latex@error\@gobble}
\makeatother

\usepackage[font=footnotesize,labelformat=simple]{subcaption}

\begin{document}

\title{Distributionally Robust Degree Optimization\\for BATS Codes}


\author{%
\IEEEauthorblockN{
Hoover~H.~F.~Yin, Jie Wang, and Sherman~S.~M.~Chow
}
\thanks{H.~H.~F.~Yin and S.~S.~M. Chow are with the Department of Information Engineering,
The Chinese University of Hong Kong.
J.~Wang is with the H.~Milton Stewart School of Industrial and Systems Engineering, Georgia Institute of Technology.
}
}

\maketitle


\begin{abstract}
Batched sparse (BATS) code is a network coding solution for multi-hop wireless networks with packet loss. Achieving a close-to-optimal rate relies on an optimal degree distribution. Technical challenges arise from the sensitivity of this distribution to the often empirically obtained rank distribution at the destination node. Specifically, if the empirical distribution overestimates the channel, BATS codes experience a significant rate degradation, leading to unstable rates across different runs and hence unpredictable transmission costs. Confronting this unresolved obstacle, we introduce a formulation for distributionally robust optimization in degree optimization. Deploying the resulting degree distribution resolves the instability of empirical rank distributions, ensuring a close-to-optimal rate, and unleashing the potential of applying BATS codes in real-world scenarios.
\end{abstract}

\section{Introduction}
Signifying a trend in the Internet of Things era, multi-hop wireless networks have gained prominence.
Signal fading, inference, and many other factors can lead to packet loss in such networks.
Meanwhile, traditional strategies for wired networks may be inadequate for wireless setups.
Contemplating a scenario where every network link has a certain packet loss rate, end-to-end retransmission is rendered inefficient as 
ensuring a packet reaches the destination node without being dropped in any network link becomes increasingly improbable with a growing number of hops.

Manifesting as a realization of network coding~\cite{linear,flow}, \emph{random linear network coding (RLNC)}~\cite{random2} adeptly attains the network capacity in many scenarios.
Instead of store-and-forward at the intermediate network nodes, store-and-compute is applied.
Named \emph{recoding}, this computational process generates random linear combinations of the received packets.
Grasping this capability, a node can then create more packets than received, thereby not conserving the packet flow.

Cognizant implementation of RLNC confronts several practical concerns.
Spearheading many chunks of packets, recoding 
them
takes a high computational cost.
Managing all these packets at the intermediate node for recoding imposes demanding computational and storage costs, particularly for typical intermediate nodes like routers or embedding devices.
Moreover, the coefficient vector attached to each packet for recording the recoding operations can be very long as each coefficient corresponds to a data packet.
Coupling these with supplementary mechanisms like encryption or authentication through homomorphic signatures~\cite{tc/ChenXYC16} further increases the costs.

Several meticulous codes, \textit{e.g.}, generation-based RLNC~\cite{chou03,crlnc}, 
overhead optimizations~\cite{deAlwis13,silva12,gligoroski15}, \textit{etc.},
contemplate enhancements to native RLNC.
One variation of RLNC is known as the \emph{batched network coding (BNC)}~\cite{silva09,heidarzadeh10,li11,tang12,mahdaviani12,tang18,yang14bats}.
In BNC, a \emph{batch} is a small set of coded packets within which 
RLNC is applied to the packets belonging to the same batch.
The formation of batches depends on an outer code such as LDPC~\cite{tang18} and generalized fountain codes~\cite{yang14bats}.

\subsection{BATS Codes and the Degree Distribution Issue}

\emph{Batched sparse (BATS) codes}~\cite{yang14bats,bats_book} are a class of BNC that have a close-to-optimal achievable rate.
As a matrix generalization of fountain codes~\cite{lubyLT},
the encoding process relies on a \emph{degree distribution}.
Nevertheless, there is no universal degree distribution for BATS codes~\cite{quasi}.
To achieve the best rate of BATS codes, we need to optimize the degree distribution according to information called the \emph{rank distribution} at the destination node.
The degree distribution optimization problem can be formulated as a linear programming problem.


\begin{figure}
	\centering
\includegraphics[width=0.5\textwidth]{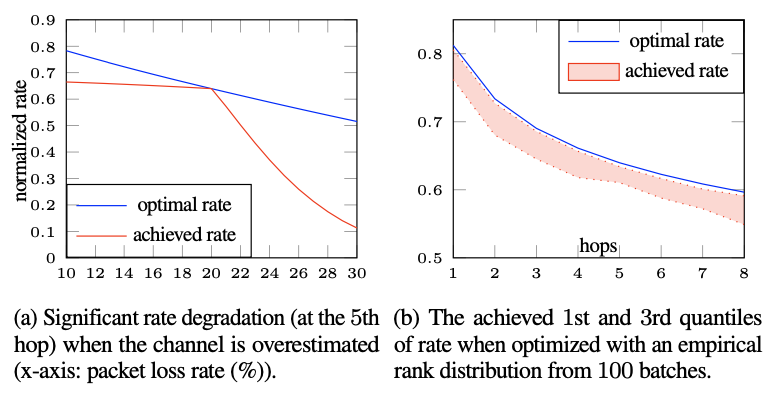}
\caption{Illustrations of significant rate degradation when overestimating the channel and its impact on rate fluctuation.}
	\label{fig:rate_drop}

\end{figure}

A critical issue of BATS codes lies in their rate sensitivity to the rank distribution provided for optimizing the degree distribution.
The rank distribution is affected by various factors, such as channel conditions and recoding policies.
It is typically derived via estimation, \textit{e.g.}, using the empirical distribution.
If the channel is underestimated, too much redundancy will be transmitted, decreasing the rate.
Overestimating the channel is lethal to the rate, as the belief propagation (BP) decoding algorithm is likely to fail.

To mitigate these issues, extra batches need to be transmitted to enable BP decoding, significantly degrading the rate.
A simple example in \Cref{fig:rate_drop} highlights its importance.
Consider a multi-hop network in which each link has the same independent packet loss rate.
Applying a BATS code with each batch consisting of $8$ coded packets, we aim to decode up to $98\%$ of the data (see precode in \Cref{sec:enc}),
\textit{i.e.}, $98\%$ of the rate in \Cref{fig:rate_drop} is capped by the capacity.

Shown in Fig.~1(a), the exact rank distribution for a $20\%$ packet loss rate is employed to optimize a degree distribution.
This degree distribution is then used across scenarios with packet loss rates per link ranging from $10\%$ to $30\%$.
The plotted BATS code rate at the $5$th hop reveals that underestimating the channel results in a rate nearly identical to that of the $20\%$-loss-rate degree distribution.
Despite a gap between the true and achieved rates, performance is stable.
However, an overestimated channel leads to a sharp drop in the rate.

Code rates achieved in Fig.~1(b) are recorded after applying a BATS code with a degree distribution optimized via an empirical rank distribution from $100$ batches, where each link has a $20\%$ packet loss rate.
The simulation is run $100$ times for each network length (the number of hops in the network).
The colored region is between the $1$st and $3$rd quantiles of the rate.
Notably, rate fluctuations across these runs indicate the instability of BATS codes' performance.
Consequently, the number of batches required for transmission is unpredictable, necessitating a huge overhead of batches to ensure reliable communication.

\subsection{Related Works}
\label{sec:related}

Most works on degree optimization for BATS codes in the literature assume the knowledge of an accurate rank distribution.
Only a few works were trying to cope with the rate degradation issue.

Optimization for all rank distributions with expectations no smaller than a threshold has been studied.
It suffices to focus solely on the rank distributions with the lowest expectation in an equivalent formulation~\cite[Ch.~6.4]{bats_book}.
This actually explains the worse case performance in the flat region of the red curve in \Cref{fig:rate_drop}, \textit{i.e.}, underestimating the channel.
For simplicity, we call it the \emph{$\mu$-universal degree optimization}.
However, the choice of threshold from the observed sample is not discussed.
That is, it cannot resolve the rate degradation issue if the threshold is too large, \textit{e.g.}, the mean of the empirical rank distribution.
More importantly, the overall rate is low as it considers too many rank distributions ``far'' away from the empirical one.

\emph{Safety margin}~\cite{fun} is another approach.
It first uses a Gaussian distribution to estimate the empirical rank distribution
and considers a worse Gaussian distribution by scaling down the mean while keeping the same variance.
At last, the worse Gaussian distribution is discretized to be the rank distribution for degree optimization.
However, it is unclear how to select a suitable scaling factor to maintain a good rate, \textit{i.e.}, it is a trade-off between the stability of the rate and the gap from the optimal rate.
There is no theoretical guarantee of the performance; say, this approach may not be robust, \textit{i.e.}, still experiencing a significant rate degradation,
when there are only a few samples to form the empirical rank distribution.
Further, the empirical rank distribution may be totally different from a Gaussian distribution in a general network, \textit{e.g.}, the variant of binomial distributions that might have two peaks~\cite{yin22mp}, so the estimation technique can be wildly inaccurate.

Quasi-universal BATS code~\cite{quasi} considers optimization for multiple rank distributions simultaneously that produces multiple degree distributions.
According to the number of batches sent, the source node switches the degree distribution to be used.
The drawback is that if the rank distribution only slightly overestimates the channel, we will eventually use some worse degree distributions to finish the transmission.
This can induce a big gap from the optimal rate.
However, this approach cannot be applied directly to the single-cast scenario we considered in this paper because we do not have multiple estimated rank distributions at the only destination node.

\subsection{Our Contribution}

Instead of mindlessly modifying the estimated rank distribution to a worse one, we consider a distributionally robust optimization (DRO) approach~(see, \textit{e.g.},~\cite{lin2022distributionally} for its comprehensive review) for degree optimization.
By considering all rank distributions in a probability ball centered at the empirical rank distribution, we optimize for the worst-case performance.
The radius of the probability ball can be calculated from the number of samples that form the empirical rank distribution to achieve an out-of-sample guarantee, \textit{i.e.}, we can avoid the rate fluctuation with a high probability.
On the other hand, when more samples are observed, the estimated rank distribution is more accurate; thus, the overall rate is higher.
This way, we can overcome the unsolved threshold and scaling factor problems in previous works.

At the heart of DRO lies the crucial task of selecting an appropriate probability metric.
We discuss the choice of probability metric for the DRO formulation and explain why Kullback-Leibler~(KL) divergence, a common metric in information theory, is unsuitable for our specific problem.
Motivated by the emerging applications of optimal transport-based DRO in recent literature~\cite{gao2023distributionally, blanchet2019quantifying, blanchet2019robust, zhang2022simple}, we construct the DRO model using the $1$-Wasserstein distance, which yields the most favorable robust rate.
We also compare the results with the total-variation distance, recently applied in some DRO literature~\cite{rahimian2019identifying,tvdro1,tvdro2}.
The results are juxtaposed with existing non-DRO approaches for a comprehensive assessment.

\section{BATS Codes}
\subsection{Notations}
For any non-negative integer $Q$, define $[Q] = \{0, 1, \ldots, Q\}$.
For any positive integer $R$, define $\llbracket R \rrbracket = \{1, 2, \ldots, R\}$.
Denote by $\mathbf{0}_S$ and $\mathbf{1}_S$ the length-$S$ zero column vector and all-one column vector respectively.
Fix a finite field $\mathbb{F}_q$ of size $q$.

\subsection{Encoding}
\label{sec:enc}

To boost the performance of BATS codes, the data to be transmitted is first encoded by an erasure code, known as a \emph{precode}, so that the data can be recovered by receiving a sufficient portion of the precoded data.
This technique was used in Raptor codes~\cite{shokRaptor} for maintaining a constant decoding complexity with respect to the data size after belief propagation decoding has been stopped.

The precoded data is then partitioned into multiple \emph{input packets}.
Each input packet is regarded as a vector over $\mathbb{F}_q$ of the same length.
An inappropriate input packet size can induce extra padding overhead.
We can apply an existing optimization~\cite{pktsize} to minimize such overhead.
Next, the encoder of a BATS code generates a sequence of batches.
A freshly generated batch by the encoder consists of $M > 0$ coded packets, where $M$ is called the \emph{batch size}.

To generate a batch, we first sample the predefined \emph{degree distribution} to obtain a \emph{degree}.
To achieve the best rate, the degree distribution must be optimized.
This degree optimization problem will be discussed in \Cref{sec:deg}.
The value of the degree, $d$, is the number of input packets to be contributed in forming the batch.
Next, we pick $d$ input packets uniformly.
Each of the $M$ coded packets in the batch is formed by taking a random linear combination of the selected input packets.
Optionally, we can further apply recoding on the $M$ coded packets to generate more packets before transmission.

The coded packets of a batch are transmitted to the next network node when they are ready.

\subsection{Recoding and Ranks of Batches}

Recoding of BATS codes is restricted to the packets belonging to the same batch.
A recoded packet of a batch is a random linear combination of the received packets of the batch.

A length-$M$ coefficient vector over $\mathbb{F}_q$ is attached to each packet to store the recoding operations.
Two packets in the same batch are said to be linearly independent if and only if their coefficient vectors are linearly independent.
At the source node, the $M$ packets in a freshly generated batch are defined as linearly independent.
This can be achieved by properly initializing the coefficient vectors~\cite{yang22pro,yin20pro}.
The number of linearly independent packets in a batch is called its \emph{rank}, which is a measure of information remaining in a batch.

The number of recoded packets to be generated and sent depends on the recoding scheme.
\emph{Basline recoding} generates the same number of recoded packets per batch regardless of the rank of the batch.
Although it is not optimal~\cite{yang14a}, it is applied in many literature for its simple implementation and analysis~\cite{variable,fun,delay,buffer,zhou17b}.
\emph{Adaptive recoding}~\cite{adaptive,scheduling,ge_adaptive,uni,bar,wang2021small,10476945} optimizes the number of recoded packets per batch according to its rank.
This way, the expected rank of the batches at the next node is maximized.
Note that these schemes affect the rank distribution of the batches arriving at the destination node.

\subsection{Decoding}

Belief propagation is the main decoding algorithm used by BATS codes.
After receiving a batch, the corresponding input packets can be recovered by solving a linear system if the rank of the batch is no smaller than the degree of the batch.
The decoded input packets are then substituted to other received batches that cannot be decoded to reduce the number of variables in them, \textit{i.e.}, reducing the effective degree of the batch.
The procedure is repeated if new input packets are being decoded.
The decoding procedure can be stopped if the portion of decoded input packets is sufficient for the precode to recover the original data.
Some other approaches, such as inactivation decoding~\cite{shokRaptor,Raptormono}, can be used to continue the decoding procedure when belief propagation decoding stops.

For simplicity, the rank distribution of the batches arriving at the destination node is called the \emph{rank distribution}.
The expected value of this rank distribution is the theoretical upper bound on the achievable rates~\cite{yang11x2}.
In other words, the ranks of the batches form a sufficient statistic for the performance of a BATS code.

\subsection{Degree Optimization Framework}
\label{sec:deg}

Maximizing the achievable rate of a BATS code requires optimizing the degree distribution contingent on the rank distribution of arriving batches at the destination node.
BATS codes for $M = 1$ degenerate into a fountain code with a universal degree distribution,
independent of the rank distribution.
However, for $M \ge 2$, there is no universal degree distribution for all rank distributions~\cite[Lem.~1]{quasi}.

In practice, the true rank distribution is uncertain, so we need to estimate it.
An inaccurate estimation can significantly degrade the performance of the BATS code, especially when the rank distribution is overestimated.
We first introduce the degree optimization problem that regards the empirical rank distribution $\mathbf{h} = (h_0, h_1, \ldots, h_M)^T$ as the true underlying rank distribution.

We aim to get a degree distribution $\boldsymbol{\Psi} = (\Psi_1, \Psi_2, \ldots, \Psi_D)^T$ 
for a fixed, maximum (integer) degree $D$
that can maximize the rate~$\theta$.
Let $\mathcal{P}$ be the probability simplex of all degree distributions, \textit{i.e.},
\begin{equation*}
	\mathcal{P} = \{ \boldsymbol{\Psi} = (\Psi_1, \Psi_2, \ldots, \Psi_D)^T \colon \|\boldsymbol{\Psi}\|_1 = 1, \boldsymbol{\Psi} \ge \mathbf{0}_D \}.
\end{equation*}
If $D$ is too small, some important degree might be ignored.
By~\cite[Thm.~6.2]{bats_book}, the rate will not improve for $D > \lceil \frac{M}{1-\eta} \rceil - 1$, where $\eta \in (0, 1)$ is the portion of the precoded data that is sufficient for the precode to recover the original data.
It is thus safe to set $D = \lceil \frac{M}{1-\eta} \rceil - 1$.

Define a vector $\pmb{\hbar} = (\hbar_0, \hbar_1, \ldots, \hbar_M)^T$ where for every $s \in [M]$, $\hbar_s$ is the probability that a batch is decodable for the first time when its degree is $s$.
This vector can be obtained by applying a linear transformation to the rank distribution.
More specifically, we can write
$\pmb{\hbar} = \mathbf{Z} \mathbf{h}$.
The matrix $\mathbf{Z}$ is an $(M+1) \times (M+1)$ upper triangular matrix.
For simplicity, we count the row and column indices of $\mathbf{Z}$ from $0$.
The entries of $\mathbf{Z}$ are
\begin{equation*}
	(\mathbf{Z})_{s,r} = \begin{cases}
		\zeta_s^r / q^{r-s} & \text{if } s \le r,\\
		0 & \text{otherwise},
	\end{cases}
\end{equation*}
where
\begin{equation*}
\zeta_r^m = \begin{cases}
 \prod_{i = 0}^{r-1} (1-q^{-m+i}) & \text{if } 0 < r \le M,\\
 1 & \text{otherwise}
\end{cases}
\end{equation*}
is the probability of an $r \times m$ totally random (every entry is i.i.d. and uniformly distributed) matrix over $\mathbb{F}_q$ is full rank.
The difference between $\pmb{\hbar}$ and $\mathbf{h}$ is negligible when the field size is large enough.
Practically, we can approximate $\mathbf{Z}$ by an identity matrix $\mathbf{I}$ when $q = 2^8$.

The necessary and sufficient condition for decoding up to $\eta \in (0, 1)$ portion of the (precoded) data can be captured by the inequality
\begin{equation} \label{eq:ori_constraint}
	\inf_{x \in (0, \eta]} (\pmb{\hbar}^T \mho(x) \boldsymbol{\Psi} + \theta \ln(1-x)) \ge 0,
\end{equation}
where for each $x$, $\mho(x)$ is an $(M+1) \times D$ matrix.
Counting the row index of $\mho(x)$ from $0$ but the column index from~$1$, its entries are
\begin{equation*}
	(\mho(x))_{r,d} = \begin{cases}
		0 & \text{if } r = 0,\\
		d & \text{if } d \le r,\\
		d I_x(d-r,r) & \text{if } d > r > 0,
	\end{cases}
\end{equation*}
where $I_x(a, b) = \frac{\int_0^x t^{a-1} (1-t)^{b-1} dt}{\int_0^1 t^{a-1} (1-t)^{b-1} dt}$ is the regularized incomplete beta function.
Inequality~\eqref{eq:ori_constraint} can be obtained via differential equation analysis~\cite{yang14bats} or tree analysis~\cite{tree}.
We omit $x = 0$ considered in previous works since it gives a trivial inequality.

The degree optimization problem is formulated in our notations as
\begin{equation*}
\begin{aligned}
\max_{\boldsymbol{\Psi} \in \mathcal{P}, \theta \in \mathbb{R}}&\quad \theta\\
\mathrm{s.t.} \quad &\quad \inf_{x \in (0, \eta]} (\mathbf{h}^T \mathbf{Z}^T \mho(x) \boldsymbol{\Psi} + \theta \ln(1-x)) \ge 0.
\end{aligned}
\end{equation*}
When $\boldsymbol{\Psi}$ is unknown, it is unclear how to find an $x$ that solves the infimum in the degree optimization problem.
In the literature~\cite{yang14bats,bats_book,quasi,sliding,expanding,unequal,variable,mao24}, 
considering discretized $x \in \mathcal{X}$,
this constraint is approximated by discretizing the interval $(0, \eta]$ into a finite set $\mathcal{X}$ with sufficiently small step size.
This constraint is expanded into $|\mathcal{X}|$ constraints.
If we know the true rank distribution $\mathbf{h}^\ast$, the rate $\tilde{\theta}$ achieved by a given degree distribution $\boldsymbol{\Psi}$ is
\begin{equation*}
	\tilde{\theta} = \inf_{x \in (0, \eta]} \frac{-(\mathbf{h}^\ast)^T \mathbf{Z}^T \mho(x) \boldsymbol{\Psi}}{\ln(1-x)}.
\end{equation*}
To align with the discretization of $x$, we can approximate $\tilde{\theta}$ by taking the minimum over $x \in \mathcal{X}$ instead.


\section{Distributionally Robust Optimization (DRO)}
The rank distribution supplied for degree optimization 
is usually obtained empirically.
Although this estimate achieves the lower bound for the minimax risk up to some constant factors~\cite{singh2019estimating}, it makes the degree optimization problem suffer from the out-of-sample performance.
In other words, we desire a degree distribution with a rate guarantee when the empirical rank distribution is inaccurate.
More specifically, we aim to maximize the rate for the worst case of all possible rank distributions having a distance smaller than some value from the empirical rank distribution, \textit{i.e.}, within a probability ball.
At the same time, we desire an out-of-sample guarantee.

In this section, we denote the empirical rank distribution, estimated from $N$ samples, by $\hat{\mathbf{h}} = (\hat{h}_0, \hat{h}_1, \ldots, \hat{h}_m)$.
Let $(\mathbf{Z}^T \mho(x))_r$ be the $r$th row of $\mathbf{Z}^T \mho(x)$ where the row index counts from $0$.
\vspace{-0.2em}
\subsection{Choice of Probability Metric}

For the probability distribution $\mathbf{h}$ and the empirical~$\hat{\mathbf{h}}$,
we choose the popular KL divergence
as the distance metric.
To achieve absolute continuity, we must set a mass of $\mathbf{h}$ to $0$ if the corresponding mass in $\hat{\mathbf{h}}$ is $0$, or otherwise the distance is defined to be $+\infty$.
This is unsuitable because the $0$ masses in the empirical distribution can be inaccurate, \textit{e.g.}, not yet received a batch of such rank.

Total variation distance is useful when no natural metric exists for entries in the support.
In our case, this can ignore whether each rank increment should be equally important.
A drawback is that the rate achieved by the total variation DRO may not be high unless there is a vast number of samples due to the consistent radius of the probability ball for various confidence levels for the out-of-sample guarantee.

Wasserstein distance is another prevalent metric in stochastic optimization.
We focus on the $1$-Wasserstein distance and use the absolute value of the difference between ranks as the metric, ensuring equal importance for each rank increment.

In this paper, we consider both the Wasserstein distance and the total variation distance to formulate the DRO for degree distribution.
We will compare their performance in \Cref{sec:numerical}.

\vspace{-0.2em}
\subsection{Wasserstein DRO}

We apply the ($1$-)Wasserstein distance defined as
\begin{equation*}
	W(\mathbf{h}, \hat{\mathbf{h}}) =
\left\{
\begin{aligned}
\min_{\substack{\gamma(r_1, r_2) \ge 0,\\ \forall r_1, r_2 \in [M]}}&~\sum_{r_1,r_2} |r_1 - r_2| \gamma(r_1, r_2)\\
\mbox{s.t.}\quad&~\sum_{r_1 \in [M]} \gamma(r_1, r_2) = \hat{\mathbf{h}},~~\forall r_2,\\
&~ \sum_{r_2 \in [M]} \gamma(r_1, r_2) = \mathbf{h},~~\forall r_1,
\end{aligned}
\right.
\end{equation*}
where $\gamma(r_1, r_2)$ is a joint probability distribution such that its marginals are $\mathbf{h}$ and $\hat{\mathbf{h}}$.
Define the ambiguity set $\mathcal{W}$ by
\begin{equation*}
	\mathcal{W} = \{\mathbf{h} \colon W(\mathbf{h}, \hat{\mathbf{h}}) \le \rho, \|\mathbf{h}\|_1 = 1, \mathbf{h} \ge \mathbf{0}_{M+1} \}.
\end{equation*}
The selection of the Wasserstein ball's radius, $\rho$, to achieve the out-of-sample guarantee will be discussed later.

In the degree optimization problem, the rank distribution $\mathbf{h}$ is involved in the constraint only.
To include the worst case, the inequality in the constraint must hold for all $\mathbf{h} \in \mathcal{W}$.
So, the Wasserstein DRO formulation for degree optimization is 
\[
\begin{aligned}
\max_{\boldsymbol{\Psi} \in \mathcal{P}, \theta \in \mathbb{R}} & \quad \theta
\\
\mbox{s.t.}\quad & \quad \theta \ln(1 - x) + \inf_{\mathbf{h} \in \mathcal{W}} \mathbf{h}^T \mathbf{Z}^T \mho(x) \boldsymbol{\Psi} \ge 0, \forall x \in \mathcal{X}.
\end{aligned}
\]
By applying the dual formulation to cope with the ambiguity set~\cite{gao2023distributionally}, we have the following theorem.

\begin{theorem} \label{thm:wass_dro}
The Wasserstein DRO formulation for degree optimization is equivalent to
\begin{equation*}
	\begin{IEEEeqnarraybox}[][c]{rCl}
		\max_{\substack{\boldsymbol{\Psi} \in \mathcal{P}, \theta \in \mathbb{R}, \boldsymbol{\lambda} \ge \mathbf{0}_{|\mathcal{X}|}\\\mathbf{s} \in \mathbb{R}^{(M+1) \times |\mathcal{X}|}}} & & \theta\\[-1.5em]
		\mathrm{s.t.}\qquad && \theta \ln(1-x) - \lambda_x \rho - \sum_{r = 0}^M \hat{h}_r s_{x,r} \ge 0, \forall x \in \mathcal{X}\\
		&& s_{x,r} \ge -(\mathbf{Z}^T \mho(x))_{r'} \boldsymbol{\Psi} - \lambda_x |r'-r|,\\
		&& \qquad\qquad\qquad\qquad \forall x \in \mathcal{X}, \forall r, r' \in [M].
	\end{IEEEeqnarraybox}
\end{equation*}
\end{theorem}

By choosing the radius $\rho$ properly, the ambiguity set $\mathcal{W}$ can be viewed as the confidence set that contains the true (but unknown) $\mathbf{h}$ with high probability, \textit{i.e.}, out-of-sample guarantee.
In the following, 
for a desired confidence level $c$, we aim to find a $\rho$ such that
\begin{equation*}
	\Pr(W(\mathbf{h}, \hat{\mathbf{h}}) \le \rho) > c.
\end{equation*}

Let $G \sim \mathcal{N}(\mathbf{0}_{M+1}, \Sigma(\mathbf{h}))$ be a multivariate Gaussian random vector whose covariance matrix $\Sigma(\mathbf{h})$ is 
\begin{equation*}
	(\Sigma(\mathbf{h}))_{r_1, r_2} = \begin{cases}
		h_{r_1} (1-h_{r_1}) & \text{if } r_1 = r_2,\\
		-h_{r_1} h_{r_2} & \text{otherwise}.
	\end{cases}
\end{equation*}
When the number of sample $N\to\infty$, it has been shown~\cite{sommerfeld2018inference} that $\sqrt{N} W(\mathbf{h}, \hat{\mathbf{h}})$ weakly converges to the distribution characterized by
\begin{equation*}
\begin{aligned}
\max_{\mathbf{u} \in \mathbb{R}^{M+1}}&\quad G^{\top} \mathbf{u}\\
\mbox{s.t.}\quad&\quad u_{r_1} - u_{r_2} \le |r_1 - r_2|, \forall r_1, r_2 \in [M].
\end{aligned}
\end{equation*}
Let $X$ denote a random variable following the above distribution.
We choose $\sqrt{N} \rho$ as the $(1-c)$-quantile of the distribution of $X$:
\begin{equation*}
	\Pr(W(\mathbf{h}, \hat{\mathbf{h}}) > \rho) \approx \Pr(X > \sqrt{N} \rho) \le 1-c.
\end{equation*}
However, $X$ involves the information of $\mathbf{h}$, which is not known precisely.
One approach to approximate $X$ by $\hat{X}$ is to replace the covariance matrix $\Sigma(\mathbf{h})$ by $\Sigma(\hat{\mathbf{h}})$.
As the probability density of $\hat{X}$ is intractable, we approximate the probability by $L$ samples of $\hat{X}$.
Then, we choose
$\rho = \hat{x}/\sqrt{N}$, 
where $\hat{x}$ be the $\lfloor L(1-c) \rfloor$-th largest value among the $L$ samples.
Usually, $L = 100$ is sufficient.

\subsection{Total Variation DRO}

The total variation distance is defined as
\begin{equation*}
	T(\mathbf{h}, \hat{\mathbf{h}}) = \frac{1}{2} \sum_{r \in [M]} |h_r - \hat{h}_r|.
\end{equation*}
%
Define the ambiguity set 
\[
\mathcal{T} = \{\mathbf{h} \colon T(\mathbf{h}, \hat{\mathbf{h}}) \le \rho, \|\mathbf{h}\|_1 = 1, \mathbf{h} \ge \mathbf{0}_{M+1} \}.
\]
Similar to the Wasserstein DRO formulation, the total variation DRO formulation for degree optimization is 
\begin{equation*}
\begin{aligned}
\max_{\boldsymbol{\Psi} \in \mathcal{P}, \theta \in \mathbb{R}}&\quad \theta\\
\mathrm{s.t.} \quad&\quad \theta \ln(1-x) + \inf_{\mathbf{h} \in \mathcal{T}} \mathbf{h}^T \mathbf{Z}^T \mho(x) \boldsymbol{\Psi} \ge 0, \forall x \in \mathcal{X}.
\end{aligned}
\end{equation*}
Following the argument similar to Theorem~\ref{thm:wass_dro}, we obtain the following finite-dimensional reformulation for total variation DRO.
\begin{theorem} \label{thm:tv_dro}
The total variation DRO formulation for degree optimization is equivalent to
\begin{equation*}
	\begin{IEEEeqnarraybox}[][c]{rCl}
		\max_{\substack{\boldsymbol{\Psi} \in \mathcal{P}, \theta \in \mathbb{R},\\\boldsymbol{\alpha}, \boldsymbol{\beta} \in \mathbb{R}^{|\mathcal{X}|}}} & \,\, & \theta\\[-.5em]
		\mathrm{s.t.}\,\, && \theta \ln(1-x) - 2 \rho \beta_x + \hat{\mathbf{h}}^T \mathbf{Z}^T \mho(x) \boldsymbol{\Psi} \ge 0, \forall x \in \mathcal{X}\\
		&& \beta_x \ge |(\mathbf{Z}^T\mho(x))_r \boldsymbol{\Psi} + \alpha_x|, \forall x \in \mathcal{X}, \forall r \in [M].
	\end{IEEEeqnarraybox}
\end{equation*}
\end{theorem}

\input{fig-eval}

\newpage
According to~\cite{canonne2020short}, for the desired confidence level $c$, we have $\Pr(T(\mathbf{h}, \hat{\mathbf{h}}) \le \rho) \ge c$
if 
\[
N \ge \frac{\max\{M+1, 2\ln(2/(1-c))\}}{\rho^2}.
\]
The radius to ensure the coverage rate of $c$ is therefore
\begin{equation*}
\rho = \sqrt{\frac{\max\{M+1, 2\ln(2/(1-c))\}}{N}}.
\end{equation*}
We have $2\ln(2/(1-c)) \ge M+1$ if and only if $c \ge 1 - 2e^{-\frac{M+1}{2}}$.
For a typical batch size, say, $M = 8$, the above inequality gives $c \gtrsim 97.778\%$.
Therefore, the maximum term is dominant by $M+1$ unless we aim at a very high confidence level.
To reduce the radius, we mainly rely on a larger $\sqrt{N}$.

\section{Numerical Evaluations}
\label{sec:numerical}

We evaluate the rate of BATS codes achieved by various approaches to obtain the degree distribution in multi-hop networks with varying numbers of hops, where all links have the same independent packet loss rate $p$.
For the optimization, we use $\eta = 0.98$, aiming to decode up to $98\%$ of the precoded data
for effective recovery of 
the original data.
This ensures that $98\%$ of the rate is capped by the capacity, thus we can see in \Cref{fig:rates} that the optimal rate at the first hop is larger than $1-p$.

To simulate practical scenarios, we obtain empirical rank distributions from $N$ batches for degree optimizations.
In each setup in \Cref{fig:rates}, we draw the curves for the $1$st and $3$rd quantiles with the region between them shaded.
The top blue curve (``optimal'') is the optimal BATS code rate achievable when the underlying rank distribution is known, serving as an upper bound on the rate.
The red region (``direct'') below is the rate by optimizing the degree distribution directly via the empirical rank distribution.
Due to the notable rate drop caused by channel overestimation, the rates in different runs differ significantly, especially when the number of samples $N$ is small, leading to a prominent red band in the plots.

To avoid channel overestimation, we scale down the mean of the empirical rank distribution by a factor of $0.9$ for both the $\mu$-universal degree distribution~\cite{bats_book} and the safety margin~\cite{fun}.
The rate for the green region (``$\mu$-universal'') is very stable, yet it is far from optimal.
The rate fluctuation for the cyan region (``safety margin'') is reduced, though its rate is lower than the $\mu$-universal rate when the number of hops is small.
Notably, the rate of safety margin drops at a slower pace than that of the $\mu$-universal one, ultimately surpassing it with an increased number of hops in the network.

\balance

For both DROs, we choose $c = 90\%$.
The rate achieved by the brown region (``total variation DRO'') highly depends on $N$.
When $N = 100$, the rate can be significantly lower than all the other approaches for the considerable radius of the probability ball to cover too many rank distributions.
When $N = 1000$, the rate surpasses that of $\mu$-universal and safety margin.
In both cases, the rate fluctuation is very small.
This suggests that not all probability metrics are conducive to maximizing the rate in DRO of degree distribution.

Lastly, the gray region (``Wasserstein DRO'') performs best among all schemes in each plot.
Positioned at the topmost part of the red region, it exhibits a minimal gap from the optimal rate while maintaining remarkable rate stability.

\section{Concluding Remarks}
The rate of BATS codes is unstable due to the use of inaccurate rank distribution for optimizing the degree distribution, which makes the transmission overhead unpredictable.
While existing approaches have mitigated the rate fluctuation, they have concurrently led to a significant reduction in the overall rate.
We presented the Wasserstein DRO solution to resolve the unstable rate issue of BATS codes while simultaneously achieving a close-to-optimal rate.
We also showed that not all probability metrics are suitable for DRO of degree distribution.
For example, the total variation DRO requires a substantial sample size to obtain a higher rate.
Future research may explore the efficiency of a more general $\phi$-divergence for DRO and compare it with Wasserstein DRO.


\newpage

\bibliographystyle{IEEEtran}
\bibliography{ref}

\clearpage

\appendices

\section{Formulations of Existing Attempts}

Here, we describe the formulations of existing attempts~\cite{bats_book,fun} to cope with the rate degradation problem, which will be applied to compare with our models in this paper.
The drawbacks and issues of these attempts were discussed in \Cref{sec:related}.

\subsection{\texorpdfstring{$\mu$-Universal Degree Optimization}{μ-Universal Degree Optimization}}

An existing attempt~\cite{bats_book} is to optimize for all rank distributions with an expectation no smaller than a threshold $\mu$.
It is proven to be equivalent~\cite{bats_book} to the problem of optimizing for all rank distributions with an expectation equal to $\mu$.
Let $\mathcal{H}_\mu$ be the probability simplex of all rank distributions with expectation $\mu$, \textit{i.e.},
\begin{equation*}
	\mathcal{H}_\mu = \{ \mathbf{h} \colon \|\mathbf{h}\|_1 = 1, \mathbf{h} \ge 0, \mathbb{E}_{X \sim \mathbf{h}}[X] = \mu \}.
\end{equation*}
The optimization problem is
\begin{equation*}
	\max_{\boldsymbol{\Psi} \in \mathcal{P}, \theta \in \mathbb{R}} \!\!\! \theta \quad \!\!\! \mathrm{s.t.} \quad \!\!\!\!\!\! \inf_{x \in \mathcal{X}} \left( \inf_{\mathbf{h} \in \mathcal{H}} \mathbf{h}^T \mathbf{Z}^T \mho(x) \boldsymbol{\Psi} + \theta \ln(1-x) \right) \ge 0.
\end{equation*}
The inner infimum is a linear program, so the vertices of $\mathcal{H}_\mu$, a finite set,
is considered~\cite{bats_book}.
For all integers $i < \mu$ and $j \ge \mu$, the rank distribution $\mathbf{h}$ is a vertex of $\mathcal{H}_\mu$~\cite{bats_book} with $h_r$ defined by: 
\begin{equation*}
	h_r = \begin{cases}
		\frac{j-\mu}{j-i} & \text{if } r = i,\\
		\frac{\mu-i}{j-i} & \text{if } r = j,\\
		0 & \text{otherwise}.
	\end{cases}
\end{equation*}
The number of vertices is upper-bounded by $\lfloor \frac{M+1}{2} \rfloor \lceil \frac{M+1}{2} \rceil$,
allowing for the expansion of infimums into multiple constraints and transforming the optimization problem into a standard linear program.

\subsection{Safety Margin}

Now, we describe the previously proposed safety margin~\cite{fun}.
The first step is to approximate the empirical rank distribution $\hat{\mathbf{h}}$ by a Gaussian distribution $\mathcal{N}(\mu, \sigma^2)$.
A common approach is the set $\mu = \mathbb{E}_{X \sim \hat{\mathbf{h}}}[X]$.
For $\sigma^2$, if we do not know the number of samples $N$ that forms $\hat{\mathbf{h}}$, then we set $\sigma^2 = \mathbb{E}_{X \sim \hat{\mathbf{h}}}[X^2] - (\mathbb{E}_{X \sim \hat{\mathbf{h}}}[X])^2$.
Otherwise, we may use the sample variance $\sigma^2 = \frac{N}{N-1} (\mathbb{E}_{X \sim \hat{\mathbf{h}}}[X^2] - (\mathbb{E}_{X \sim \hat{\mathbf{h}}}[X])^2)$.

After that, we scale the mean by a factor $s \in (0, 1)$ and obtain the Gaussian distribution $\mathcal{N}(s\mu, \sigma^2)$.
By discretizing this Gaussian distribution, we obtain the rank distribution $\mathbf{h}$ for the degree optimization problem.
Let $X \sim \mathcal{N}(s\mu, \sigma^2)$.
One possible discretization is
\begin{equation*}
	h_r = \begin{cases}
		\Pr(r - 0.5 \le X < r + 0.5 ) & \text{if } 1 \le r < M,\\
		\Pr(X \ge M - 0.5) & \text{if } r = M,\\
		\Pr(X < 0.5) & \text{if } r = 0.
	\end{cases}
\end{equation*}

\section{Proof of \Cref{thm:wass_dro}}

To obtain a tractable reformulation, we need to transform the infimum in the constraint into a supremum so that the supremum can be merged with the maximum of the whole optimization problem.
Now, fix an $x \in \mathcal{X}$ and we focus on the infimum part
\begin{equation*}
	\inf_{\mathbf{h} \in \mathcal{W}} \mathbf{h}^T \mathbf{Z}^T \mho(x) \boldsymbol{\Psi}.
\end{equation*}
By applying the dual formulation~\cite{gao2023distributionally}, we obtain
\begin{equation*}
	\sup_{\lambda_x \ge 0} \left( -\lambda_x \rho - \frac{1}{N} \sum_{j = 1}^N \sup_{r' \in [M]} (-(\mathbf{Z}^T \mho(x))_{r'} \boldsymbol{\Psi} - \lambda_x |r'-\hat{r}_j|) \right)\!,
\end{equation*}
where $N$ is the number of samples, and $\hat{r}_j$ is the $j$-th sample, \textit{i.e.}, the rank of the $j$-th batch arriving at the destination node.
By counting the occurrences of the ranks of the samples, we have
\begin{equation*}
	\sup_{\lambda_x \ge 0} \left( -\lambda_x \rho - \sum_{r = 0}^M \hat{h}_r \sup_{r' \in [M]} (-(\mathbf{Z}^T \mho(x))_{r'} \boldsymbol{\Psi} - \lambda_x |r'-r|) \right).
\end{equation*}
The outer supremum can then be merged with the maximum of the whole optimization problem.
By considering all $x \in \mathcal{X}$, we have
\begin{multline*}
	\max_{\boldsymbol{\Psi} \in \mathcal{P}, \theta \in \mathbb{R}, \boldsymbol{\lambda} \ge \mathbf{0}_{|\mathcal{X}|}} \theta\\
		\mathrm{s.t.} \quad -\sum_{r = 0}^M \hat{h}_r \sup_{r' \in [M]} (-(\mathbf{Z}^T \mho(x))_{r'} \boldsymbol{\Psi} - \lambda_x |r'-r|)\\
		+\,\theta \ln(1-x) - \lambda_x \rho \ge 0, \forall x \in \mathcal{X},
\end{multline*}
This formulation is equivalent to
\begin{equation*}
	\begin{IEEEeqnarraybox}[][c]{rCl}
		\max_{\substack{\boldsymbol{\Psi} \in \mathcal{P}, \theta \in \mathbb{R}, \boldsymbol{\lambda} \ge \mathbf{0}_{|\mathcal{X}|}\\\mathbf{s} \in \mathbb{R}^{(M+1) \times |\mathcal{X}|}}} & & \theta\\
		\mathrm{s.t.}\,\,\,\,\,\, & & \theta \ln(1-x) - \lambda_x \rho - \sum_{r = 0}^M \hat{h}_r s_{x,r} \ge 0, \forall x \in \mathcal{X}\\
		&& s_{x,r} \ge -(\mathbf{Z}^T \mho(x))_{r'} \boldsymbol{\Psi} - \lambda_x |r'-r|,\\
		&& \qquad\qquad\qquad\qquad \forall x \in \mathcal{X}, \forall r, r' \in [M].
	\end{IEEEeqnarraybox}
\end{equation*}

The proof is completed.

\section{Proof of \Cref{thm:tv_dro}}

Similar to the proof of \Cref{thm:wass_dro}, we want to transform the infimum in the constraint into a supremum.
Fix an $x \in \mathcal{X}$ and consider the infimum part
\begin{equation*}
	\inf_{\mathbf{h} \in \mathcal{T}} \mathbf{h}^T \mathbf{Z}^T \mho(x) \boldsymbol{\Psi}.
\end{equation*}
This problem is equivalent to 
\begin{equation*}
	\begin{IEEEeqnarraybox}[][c]{rCl}
		\inf_{\mathbf{h}, \mathbf{s} \in \mathbb{R}^{M+1}} & \quad & \mathbf{h}^T \mathbf{Z}^T \mho(x) \boldsymbol{\Psi}\\
		\mathrm{s.t.} && \frac{1}{2} \sum_{r = 0}^M s_r = \rho, \quad \sum_{r = 0}^M h_r = 1\\
		&& -s_r \le h_r - \hat{h}_r \le s_r, \quad h_r \ge 0, \quad \forall r \in [M].
	\end{IEEEeqnarraybox}
\end{equation*}
The Lagrangian function is
\begin{IEEEeqnarray*}{Cl}
	& \mathbf{h}^T \mathbf{Z}^T \mho(x) \boldsymbol{\Psi} - \sum_{r = 0}^M \lambda_{x, r} h_r + \alpha_x \left( \sum_{r = 0}^M h_r - 1 \right)\\
	& +\,\beta_x \left( \frac{1}{2} \sum_{r = 0}^M s_r - \rho \right) + \sum_{r = 0}^M \mu_{x, r} ( h_r - \hat{h}_r - s_r )\\
	& -\,\sum_{r = 0}^M \nu_{x, r} ( h_r - \hat{h}_r + s_r )\\
	= & \sum_{r = 0}^M ( (\mathbf{Z}^T \mho(x))_r \boldsymbol{\Psi} - \lambda_{x, r} + \alpha_x + \mu_{x, r} - \nu_{x, r} ) h_r\\
	& +\,\sum_{r = 0}^M \left( \frac{1}{2} \beta_x - \mu_{x, r} - \nu_{x, r} \right) s_r - \alpha_x - \beta_x \rho\\
	& +\,\sum_{r = 0}^M \hat{h}_r ( \nu_{x, r} - \mu_{x, r} ).
\end{IEEEeqnarray*}
That is, the dual formulation is
\begin{equation*}
	\begin{IEEEeqnarraybox}[][c]{rCl}
		\sup_{\substack{\boldsymbol{\lambda}_x, \boldsymbol{\mu}_x, \boldsymbol{\nu}_x \ge \mathbf{0}_{M+1}\\\alpha_x, \beta_x \in \mathbb{R}}} && - \alpha_x - \beta_x \rho + \sum_{r = 0}^M \hat{h}_r ( \nu_{x, r} - \mu_{x, r} )\\
		\mathrm{s.t.}\,\,\,\,\,\,\,\,\,\,\,\,\,\,\,\,\,\,\,\,\,\,\,\,\,\,\, && \!\!\!\!\!\!\!\!\!\!\!\!\!\!\!\!\!\!\!\!\!\!\!\!(\mathbf{Z}^T \mho(x))_r \boldsymbol{\Psi} - \lambda_{x, r} + \alpha_x + \mu_{x, r} - \nu_{x, r} = 0, \forall r \in [M]\\
		&& \frac{1}{2} \beta_x - \mu_{x, r} - \nu_{x, r} = 0, \forall r \in [M].
	\end{IEEEeqnarraybox}
\end{equation*}

We now simplify the dual formulation.
The first constraint gives 
\begin{equation*}
	\nu_{x, r} - \mu_{x, r} = (\mathbf{Z}^T \mho(x))_r \boldsymbol{\Psi} - \lambda_{x, r} + \alpha_x, \forall r \in [M],
\end{equation*}
which we can substitute for the objective.
Together with the second constraint, we have
\begin{IEEEeqnarray*}{rCl}
	\mu_{x, r} & = & \frac{1}{2} \left( (\mathbf{Z}^T \mho(x))_r \boldsymbol{\Psi} - \lambda_{x, r} + \alpha_x + \frac{1}{2} \beta_x \right) \ge 0,\\
	\nu_{x, r} & = & \frac{1}{2} \left( - (\mathbf{Z}^T \mho(x))_r \boldsymbol{\Psi} + \lambda_{x, r} - \alpha_x + \frac{1}{2} \beta_x \right) \ge 0
\end{IEEEeqnarray*}
for all $r \in [M]$.
This means that
\begin{equation*}
	\frac{1}{2} \beta_x \ge | (\mathbf{Z}^T \mho(x))_r \boldsymbol{\Psi} - \lambda_{x, r} + \alpha_x |, \forall r \in [M].
\end{equation*}
Equivalently, we have
\begin{equation*}
	\beta_x \ge 2 \sup_{r \in [M]} | (\mathbf{Z}^T \mho(x))_r \boldsymbol{\Psi} - \lambda_{x, r} + \alpha_x | \ge 0.
\end{equation*}
Therefore, the dual formulation becomes
\begin{equation*}
	\begin{IEEEeqnarraybox}[][c]{rCl}
		\sup_{\substack{\boldsymbol{\lambda}_x \ge \mathbf{0}_{M+1}\\\alpha_x, \beta_x \in \mathbb{R}}} & \quad & - \beta_x \rho + \hat{\boldsymbol{h}}^T ( \mathbf{Z}^T \mho(x) \boldsymbol{\Psi} - \boldsymbol{\lambda}_x )\\
		\mathrm{s.t.} && \beta_x \ge 2 \sup_{r \in [M]} | (\mathbf{Z}^T \mho(x))_r \boldsymbol{\Psi} - \lambda_{x, r} + \alpha_x |.
	\end{IEEEeqnarraybox}
\end{equation*}

Note that $\lambda_{x, r} \ge 0$.
If $(\mathbf{Z}^T \mho(x))_r \boldsymbol{\Psi} - \lambda_{x, r} + \alpha_x \ge 0$, then we need $\boldsymbol{\lambda}_x = \mathbf{0}_{M+1}$ to achieve the supreme in the constraint.
That is, the problem becomes
\begin{equation*}
	\begin{IEEEeqnarraybox}[][c]{rCl}
		\sup_{\substack{\boldsymbol{\lambda}_x \ge \mathbf{0}_{M+1}\\\alpha_x, \beta_x \in \mathbb{R}}} & \quad & - \beta_x \rho + \hat{\boldsymbol{h}}^T ( \mathbf{Z}^T \mho(x) \boldsymbol{\Psi} - \boldsymbol{\lambda}_x )\\
		\mathrm{s.t.} && \beta_x \ge 2 \sup_{r \in [M]} \max\{ (\mathbf{Z}^T \mho(x))_r \boldsymbol{\Psi} + \alpha_x,\\
		&& \qquad\qquad\qquad \lambda_{x, r} - (\mathbf{Z}^T \mho(x))_r \boldsymbol{\Psi} - \alpha_x \}.
	\end{IEEEeqnarraybox}
\end{equation*}

In the objective, $\beta_x \ge 0$ must be as small as possible to achieve the optimal value, thus we must have $\boldsymbol{\lambda}_x = \mathbf{0}_{M+1}$ again, so we can further simplify the dual formulation into
\begin{equation*}
	\sup_{\alpha_x \in \mathbb{R}} \left( - 2 \rho \sup_{r \in [M]} | (\mathbf{Z}^T \mho(x))_r \boldsymbol{\Psi} + \alpha_x | + \hat{\boldsymbol{h}}^T \mathbf{Z}^T \mho(x) \boldsymbol{\Psi} \right).
\end{equation*}

The last step is to merge the supremum with the maximum of the whole optimization problem.
This gives
\begin{equation*}
	\begin{IEEEeqnarraybox}[][c]{rCl}
		\max_{\substack{\boldsymbol{\Psi} \in \mathcal{P}, \theta \in \mathbb{R}\\\boldsymbol{\alpha}, \boldsymbol{\beta} \in \mathbb{R}^{|\mathcal{X}|}}} & \,\, & \theta\\
		\mathrm{s.t.}\,\, && \theta \ln(1-x) - 2 \rho \beta_x + \hat{\mathbf{h}}^T \mathbf{Z}^T \mho(x) \boldsymbol{\Psi} \ge 0, \forall x \in \mathcal{X}\\
		&& \beta_x \ge |(\mathbf{Z}^T\mho(x))_r \boldsymbol{\Psi} + \alpha_x|, \forall x \in \mathcal{X}, \forall r \in [M].
	\end{IEEEeqnarraybox}
\end{equation*}

The proof is completed.

\end{document}